**Observation and Modeling of the Solar Transition Region: II. Solutions of the Quasi-Static Loop Model**

Hakeem M. Oluseyi
Department of Physics
Center for Space Sciences and Astrophysics
Stanford University
Stanford, CA 94305-4060 USA
(650) 723 - 3216
Hakeem@Banneker.Stanford.edu

A. B. C. Walker, II
Departments of Physics and of Applied Physics
Stanford University
Stanford, CA 94305-4060 USA
(650) 723 - 1486
Walker@Banneker.Stanford.edu

David I. Santiago
Departments of Physics
Stanford University
Stanford, CA 94305-4060 USA
(650) 723 - 1775
David@spacetime.stanford.edu

Richard B. Hoover
Space Sciences Laboratory
NASA Marshall Space Flight Center
Huntsville, AL 35812 USA
(256) 544 - 7617
Richard.Hoover@msfc.nasa.gov

Troy W. Barbee, Jr.
Chemistry and Materials Science Department
Lawrence Livermore National Laboratory
Livermore, CA 94550 USA
(925) 423 - 7796
barbee2@llnl.gov

# ABSTRACT

In the present work we undertake a study of the quasi-static loop model and the observational consequences of the various solutions found. We obtain the most general solutions consistent with certain initial conditions. Great care is exercised in choosing these conditions to be physically plausible (motivated by observations). We show that the assumptions of previous quasi-static loop models, such as the models of Rosner, Tucker and Vaiana (1978) and Veseckey, Antiochos and Underwood (1979), are not necessarily valid for small loops at transition region temperatures. We find three general classes of solutions for the quasi-static loop model, which we denote, radiation dominated loops, conduction dominated loops and classical loops. These solutions are then compared with observations. Departures from the classical scaling law of RTV are found for the solutions obtained. It is shown that loops of the type that we model here can make a significant contribution to lower transition region emission via thermal conduction from the upper transition region.

## 1. INTRODUCTION

High-resolution soft X-Ray and EUV observations of the Sun have revealed that the basic structural form for the hot solar atmosphere is the plasma loop. This structuring appears to be determined by the local magnetic field. In lines emitted from ions with temperatures ranging from the lower transition region (20 000 K) through the corona ($10^6$ K), loops have been directly observed, or inferred to exist, above photospheric magnetic bipoles (Prés & Phillips 1999; Kankelborg *et al*. 1996, 1997; Falconer *et al*. 1996, 1998; Dowdy 1993; Fontenla *et al*. 1989; Mariska 1986; Bonnet 1980) for a wide range of size scales. Recent observations with high temporal and spatial resolution show that many of these loop structures are transient, dynamically evolving structures (see for example, Kjeldseth-Moe & Brekke 1998, Yun *et al*. 1998, Wang *et al*. 1997, Korendyke *et al*. 1995, Strong 1994). Nevertheless, a large number of loop structures have been observed to persist for periods of time that are long, compared to radiative and conductive cooling time-scales for a fully ionized plasma, without undergoing any dramatic changes in either their luminosities or structures (see for example, Craig & McClymont 1981; Veseckey *et al*. 1979, Gerassimenko *et al*. 1978). For such loops, quasi-static loop models such as those developed by Rosner, Tucker and Vaiana (1978) (hereafter RTV) and Veseckey, Antiochos and Underwood (1979) (hereafter VAU) are appropriate.

The quasi-static loop model has been applied to observations of large loops in the corona by Waljeski, Dere and Moses (1992), Klimchuck & Porter (1995), Porter & Klimchuk (1995), and Kano & Tsuneta (1995, 1996). The model has also been applied to observations of X-Ray bright points (XBPs) in the corona by Kankelborg *et al*. (1996, 1997), and to observations of the upper transition region by Oluseyi *et al*. (1999). In the two latter studies the lower transition region was not explicitly included in the models, abandoning the boundary conditions chosen in the RTV and VAU models which required the conductive energy flux to disappear at the base of the transition region. Instead, Kankelborg *et al*. (1996, 1997) took advantage of the high spatial resolution and broad spectral coverage of their observations by the Multi-Spectral Solar Telescope Array (MSSTA) (Allen *et al*. 1993, Lindblom *et al*. 1991), and chose their boundary conditions by specifying the XBPs' basal temperatures to be consistent with temperature of formation of the strong solar emission line, H Lyα. Kankelborg *et al*. chose their second boundary condition by requiring the XBPs' basal temperature gradients (and, hence, conductive fluxes) to be consistent with the local H Lyα flux simultaneously observed by the MSSTA. Oluseyi *et al*. (1999) used observations by the MSSTA pre-cursor which did not have as broad a spectral coverage, and hence, no simultaneous H Lyα spectroheliogram. Nevertheless, the conditions at the bases of the Oluseyi models were chosen to be consistent with the area averaged H Lyα flux observed by the MSSTA and by other observers. This approach was motivated by the work of Fontenla, Avrett, and Loeser (1990, 1991, 1993) (hereafter FAL) who showed that below $10^5$ K ambipolar diffusion becomes important in the energy balance of the lower transition region. The FAL findings also show that most of the energy conducted below the $10^5$ K isotherm will be radiatively dissipated, primarily by H Lyα radiation. This is because the hydrogen ionization energy flow due to ambipolar diffusion enables the lower transition region to increase its radiative losses in H Lyα in response to high temperature gradients. Since the observed H Lyα losses provide an indirect measurement of the temperature gradient at the temperature of formation of the H Lyα line, we believe this method to be more appropriate than ad hoc boundary conditions.

An essential feature of the analytical quasi-static loop model derived by RTV is the classical scaling law, which relates a coronal loop's peak temperature, $T_m$, with the product of its pressure and length, $T_m \propto (PL)^{\alpha}$, with $\alpha = 1/3$ in the RTV derivation. This scaling law is derived based on the assumption that radiation and conduction are the only heat loss mechanisms for a loop down to chromospheric temperatures (assumed to be 20 000 K by RTV and 30 000 K by VAU). Based on this assumption VAU argue the validity of the boundary condition RTV choose at the base of the transition region (*i.e.* that the conductive flux vanishes there). VAU also argue that this assumption leads to the conclusion that the ratio of the radiative losses to the conductive losses for the coronal portion of a loop should be of order unity. With regard to the assumption used in the VAU argument, VAU states "what is needed to invalidate this argument is a heat sink other than radiation." We believe that the work of FAL points to the existence of such a heat sink, and hence, the scaling law of RTV need not strictly hold. We note that another questionable assumption of the RTV & VAU models, that LTE conditions prevail from the corona to the chromosphere, also motivates us to revisit the conclusions of their model. In this paper we explore the solutions of the quasi-static loop model of RTV and VAU without the restrictions of their models, and explore the observational consequences of the solutions found.

## 2. ANALYSIS

The current study is motivated by the work of Oluseyi *et al.* (1999) (hereafter Paper I), where we applied the quasi-static loop model to unresolved structures observed in a 171 Å – 175 Å bandpass by the MSSTA. The observed emission was modeled as a distribution of small constant cross-section loops (~ 15″ total length with 1″ diameter as observed from earth), with maximum temperatures in the upper transition region, $5 \times 10^5$ K < $T_{max}$ < $9 \times 10^5$ K, to agree with MSSTA observations. The loops' basal temperatures were set at $10^5$

K, below which the assumptions of LTE are not necessarily valid. Below $10^5$ K we rely upon the model of FAL to describe the conditions in the loop. This assumption is the key difference between our models and those of RTV and VAU. If this assumption does not hold then the results we derive may not be valid. On the other hand, the models of FAL seems to be supported by observations. We note that the heat flux at $10^5$ K in the FAL models shows good agreement with the observed H Lyα flux. Furthermore, FAL placed the chromosphere at ~ 8000 K in their models and significant heat fluxes (due to ambipolar diffusion) are present in their models down to chromospheric temperatures. At the base of their LTR model there is a residual heat flux on the order of 100 ergs $cm^{-2}$ $s^{-1}$. The FAL models also reproduce details of the emission profiles for hydrogen and helium at transition region temperatures, giving us confidence in their approach. The conditions we have chosen are very different from the conditions chosen by previous authors. The name "lukewarm loops" was coined for the loops whose conditions are described by the Oluseyi *et al*. models.

In previous models the loops' geometries and temperatures were chosen to reflect the observations of a typical coronal loop observed in quiet or active regions by grazing incidence telescopes. These instruments are typically insensitive to radiation with photon energies less than ~ 300 eV (~ 40 Å) due to the use of thick filters, and have resolutions of ~ 5″ or more; hence, the loops they observe are typically hotter ($T > 10^6$ K) and larger (total length > 30″) than the loops we model. In addition, previous modelers assumed an energy balance equation with radiation and conduction as the only energy loss mechanisms, was applicable from the corona to the chromosphere. Hence, they chose boundary conditions at the bases of their models to reflect chromospheric conditions. The basal temperatures of their models were set at ~ 2 or $3 \times 10^4$ K. Furthermore, it was argued that the conductive flux disappears at the chromosphere, so as to be consistent with a static chromospheric model in which radiation is the only energy loss mechanism. These choices of the physical conditions of the loops led directly to the scaling laws of RTV. In the next section we will show that the assumptions of the RTV scaling law do not apply to the lukewarm loops described in Paper I.

One reason for seeking scaling laws such as that derived by RTV, is that they allow the mathematical convenience that measurement of just two physical parameters of the system (for example $T_m$ and $L$), allows one to uniquely determine the third parameter (in this case the loop's pressure, $p$). We note that if the scaling laws are not strictly valid, then specifying two of the physical parameters of a loop do not uniquely determine the third under all circumstances. In other words, it would be possible to have loops with very similar lengths and peak temperatures, but with different pressures, and hence, densities.

### *2.1 The Loop Equation*

The classical one dimensional loop model of RTV and VAU balances a constant energy input with radiative and conductive losses,

$$\nabla \bullet F_c = \frac{1}{A}\frac{d}{ds}\left(AF_c\right) = \varepsilon - n_e^2(T)\Lambda(T) \tag{1}$$

where $F_c$ is the conductive thermal energy flux through a unit area along the loop, $\varepsilon$ is the energy input per unit volume, $n_e^2\Lambda(T)$ is the radiative energy loss per cubic centimeter and $A$ represents the cross-sectional area of the loop. For the portion of the loop with $T_e \geq 10^5$ K, the classical Spitzer conductivity for fully ionized plasmas is appropriate,

$$F_c = -\kappa T^{5/2} \frac{dT}{ds} \tag{2}$$

where $\kappa \sim 10^{-6}$. However, below $10^5$ K the total particle heat flux $F_H = F_c + F_A$, where $F_A$ represents the energy flux due to ambipolar diffusion, is more appropriate. The radiative loss function $\Lambda(T)$ (in ergs cm$^3$ s$^{-1}$) was analytically approximated by a sequence of power laws of the form

$$\Lambda(T) = \Lambda_s(T/T_s)^M, \tag{3}$$

joined continuously. We note that the form for $\Lambda(T)$ given in (3) is based on the assumption of LTE, which may not be strictly valid below $10^5$ K. If we assume that the loops are small compared to the gravitational scale-height, we may neglect gravity and our equation of state becomes,

$$p = 2n_e kT = \text{constant} \tag{4}$$

Using equations (3) and (4), we may rewrite equation (1) as

$$\left(F_c / \kappa T^{5/2}\right)\left(dF_c / dT\right) = \left(p^2 / 4k^2 T^2\right)\Lambda(T) - \varepsilon \tag{5}$$

Equation (5) can be integrated to obtain the thermal conductive flux as a function of temperature:

$$F_c^2(T) - F_c^2(T_0) = \left(\kappa p^2 / 2k^2\right)\int_{T_0}^{T} dT' \, T'^{1/2} \, \Lambda(T') - 2\kappa\varepsilon\int_{T_0}^{T} dT' \, T'^{5/2} \tag{6}$$

Equation (6) may be rewritten as,

$$F_c(T) = \sqrt{F_c^2(T_0) + f_R - f_H} \, . \tag{7}$$

where,

$$f_R \equiv \left(\kappa p^2 / 2k^2\right)\int dT' \, T'^{1/2} \, \Lambda(T') \tag{8}$$

and,

$$f_H \equiv 2\kappa\varepsilon\int_{T_0}^{T} dT' \, T'^{5/2} \, . \tag{9}$$

Equation (2) may be integrated to obtain:

$$s(T) - s(T_0) = \int_{T_0}^{T} \frac{T'^{5/2} \, dT'}{F_c(T')} \, . \tag{10}$$

$s_{max}$ is defined by $s_{max} \equiv s(T_{max})$, and $T_{max}$ is determined from $F_c(T_{max}) = 0$. In order to derive their scaling law, RTV made several simplifying assumptions. Their assumptions are not generally true, and, in particular, do not hold for the lukewarm loops. RTV imposed certain conditions on the form of Λ(T) and the value of ε. Also, they neglected $F_c^2(T_0)$ [$< 10^4$ at the base of their model] in comparison to $f_R$ and $f_H$ ($\sim 10^{10}$). As previously mentioned, $T_{max}$ is determined by the condition that the conductive flux vanish. Hence, near the top of the loop $F_c^2(T_0)$ becomes important as it must be exactly cancelled by $f_R$ minus $f_H$. Therefore, the scaling law should be roughly true for relatively large loops in which the region near the top of the loop where $F_c^2(T_0)$ becomes important, does not contribute much, compared to the rest of the loop. That these assumptions fail for small lukewarm loops may be seen from the following. In small lukewarm loops the conductive flux at the base is ~ $10^5$ and hence, $F_c^{\,2}(T_0)$ will be comparable to $f_R$ and $f_H$ throughout the loop, and may be even more important than $f_R$ and $f_H$ near the base.

The assumptions of RTV lead to other results that are necessary to derive their scaling law, which are not necessarily valid for the lukewarm loops. For instance, the previous models require the heat flux to disappear at $2 - 3 \times 10^4$ K. The results of FAL show that this assumption is not necessarily true. In fact, a residual heat flux remains even at 8000 K. Also, it is necessary to have $T_{max} >> T_0$ to derive the exact form of the RTV scaling law. For lukewarm loops, however, $T_{max}$ is at most $9T_0$. Another conclusion drawn of previous loop models based on the RTV assumptions is that the lower transition region portion of a coronal loop encompasses such a small portion of a loop's volume that one may neglect this portion of the loop. In fact VAU state, "…if the coronal losses are negligible, the radiation from cooler material (e.g., the transition region), is even less significant… The key point is that, although cooler material is a more effective radiator, there is much less of it..." However, for the lukewarm loops the volume of material lying below $10^5$ K is a much larger percentage of a loop's volume because the loops are so small. If we assume that the lower transition region of our loops is consistent with the FAL models, we find that the lower transition region accounts for ~ 10 % of a loop's volume. In this scenario radiative losses for the lower transition region are non-negligible, as we shall show later.

### *2.2 Computational Method*

The two boundary conditions of our model are the temperature at the base of the loop, $T_0 = 10^5$ K, and the temperature gradient (and hence, the conductive flux, $F_c(T_0)$) at the base of the loop. To compute solutions for our lukewarm loop model we first choose our boundary conditions at the base of the loop ($T_0$) and $F_c(T_0)$, and find, for a particular choice of the parameters $\varepsilon$ and $n_e(T_0)$, the position s along the loop where dT/ds = 0. Equation (9) is then integrated from the base of the loop, where $T = 10^5$ K, to the apex, where dT/ds = 0, using a fifth-order open Romberg method. The result is a determination of the loop half-length, L, the maximum temperature $T_{max}$, the conductive flux, $F_c$ and the temperature and density profiles of the loop. The model solutions were constrained by requiring the lengths of the loops to fall within our chosen size parameters, and requiring the loops' peak temperatures to lie within the upper transition region, 500 000 K < $T_e$ < 900 000 K.

For clarity, we note once again that the value of $F_c(T_0)$ was chosen to match the observed H Lyα flux and the heat flux derived in the FAL models. Based on the FAL models, $10^4 \leq F_c(T_0) \leq 10^6$; these values are also consistent with observations of the H Lyα flux. Hence, we avoid obtaining unphysical conditions at the base of the transition region that are inconsistent with the chromosphere, as would occur if we simply assumed that the transition region were able to accommodate any arbitrary flux. We also note that our choice of $n_e(T_0)$ is not arbitrary either. The values chosen for $n_e(T_0)$ are constrained by observations, as well as by keeping them consistent with the values chosen by FAL. The values of the loops' cross-sectional areas, $A$, while somewhat arbitrary, are also constrained by observation. We note that the thermal structure of the loops is not dependent on the loops' diameters.

## 3. Results and Discussion

In Paper I the models were chosen by comparing them to observations obtained by the MSSTA pre-cursor in 1987. In that study we restricted ourselves to models where the radiative flux from the loop was approximately equal to the conductive flux at the loop's base, thus guaranteeing that the solutions were consistent with the RTV scaling laws. In this study we consider a wider variety of solutions. For the loops we model, it turns out that there are three different domains of the parameter space that may yield good matches to the data. Loops in the first domain we denote *radiation dominated* loops. These loops are characterized by a large volumetric energy input, $\varepsilon$, large pressures, and relatively small temperature gradients (as compared to the other classes of models). Loops in this domain may dissipate over 90 % of their input energy via radiation. Loops in the second domain we denote *conduction dominated* loops. These loops are characterized by small volumetric energy input, low pressure and large temperature gradients. These loops may dissipate two-thirds, or more, of their energy via thermal conduction to cooler plasmas at the loops' footpoints. The third domain we denote *classical* loops. These loops have moderate energy inputs, pressures and temperature gradients. They dissipate their energy roughly equally by radiation and conduction. These classifications are chosen based on observational considerations that we will describe in the next section.

### *3.1 Comparison with Observations.*

Table 1 lists the peak temperatures, $T_{max}$, volumetric heating rates, $\varepsilon$, basal electron densities, $n_{eo}$, apex electron densities, $n_{eL}$, half-lengths, $L$, basal conductive heat fluxes, $F_{c0}$, radiative losses, $F_R$, and mechanical energy input flux, $\varepsilon L$, for each type of loop. Loops for three temperature regimes (550 000 K – 600 000 K, 700 000 K – 750 000 K, and ~ 850 000 K – 900 000 K) are shown. The models are denoted R for radiation dominated, C for conduction dominated and CL for classical.

Figure 2 shows the temperature and density profiles for representative conductive, classical, and radiative models. Figure 3 shows the differential emission measure profiles for our representative models. In order to match the models calculated above to our observational data, we calculate the emission these models would produce in the bandpass of our telescope. Using the Landini and Fossi (1990) line emissivity calculations, we have calculated the luminosities (ergs $s^{-1}$) each model would produce in the six strongest lines in our bandpass for a 1″ cross-section loop (see Table 2): Fe IX (λ 171.07 Å), Fe X (λ 174.51 Å), O V (λ 172.17 Å), O VI (λ 173.03 Å), Ne IV (λ 172.60 Å), and Ne V (λ 173.93 Å).

Using model CL2 as an example, we calculate the disk coverage necessary to produce our observed emission. In Paper I we measured the quiet sun flux in our 171 Å – 175 Å bandpass to be ~ 2.0 – 2.5 × $10^3$ ergs $cm^{-2}$ $s^{-1}$ at the sun. In Table 4 we have calculated the emission in our bandpass that would be produced from model CL2 to be 6.88 × $10^{19}$ ergs/sec. Dividing by the maximum projected area of the loop (2.25 × $10^{16}$ $cm^2$) gives us 3.06 × $10^3$ ergs $cm^{-2}$ $s^{-1}$. The ratio of our measured flux, ~ 2.0 – 2.5 × $10^3$ ergs $cm^{-2}$ $s^{-1}$, and our calculated flux, 3.06 × $10^3$ ergs $cm^{-2}$ $s^{-1}$, implies that this type of model would match our observed data with ~ 65 – 80 % disk coverage. Table 3 summarizes the data for each of our representative models.

From Table 3 we see that none of the conductive models are able to yield enough flux to match our data. Among the classical models, models CL2 and CL3 with 700 000 K ≤ $T_e$ ≤ 900 000 K may yield good matches to our observed data with reasonable disk coverages, whereas model CL1 does not yield a radiative flux consistent with our observations. While all of the radiative models yield fluxes whose magnitude may satisfy our observed flux, the calculated coverages (~ 10%) for models R2 and R3 are less than the observed coverage of the unresolved quiet sun emission. We note that Kankelborg *et al*. (1996, 1997) observed both conductive and classical type loops in the 26 XBPs they modeled conductive ($0.25 \leq F_R/F_c \leq 0.70$). More recently, Prés & Phillips (1999) observed 6 XBPs that were all strongly conductive ($0.02 \leq F_R/F_c \leq 0.1$) while Aschwanden *et al*. (1998) observed 30 active region loop segments and found them all to be strongly radiative ($F_R/F_c > 100$). These observations show that the conclusion of the VAU argument mentioned previously, that the ratio of the radiative to the conductive losses for the coronal portion of a loop is of order unity, does not strictly hold.

Several previous observers have compared the results of their models and observations of coronal loops to the coronal loop scaling law of Rosner, Tucker and Vaiana (1978) (Garcia 1998; Kankelborg *et al*. 1997, 1996; Kano & Tsuneta 1996, 1995; Porter & Klimchuk 1995). Figure 4 show the results of plotting $T_{max}$ vs ($PL$) for our representative models along with the results found for the 26 XBPs of Kankelborg *et al*. (1997, 1996). We note that several previous studies have found variation in the value of $\alpha$ inferred from observation (Kankelborg *et al*. 1997, 1996; Kano & Tsuneta 1996, 1995; Porter & Klimchuk 1995; FAL; Roberts & Frankenthal 1980). For example, Kankelborg *et al*. derive a value of $\alpha \sim 1/2.5$, Kano & Tsuneta derive a value of $\alpha \sim 1/5.1$, while RTV have $\alpha = 1/3$. Interestingly enough, we showed in Paper I that when our classical models are plotted with the XBPs of Kankelborg *et al*. (1997, 1996) we obtain a power law index of $\alpha \sim 1/5.2$ in close agreement with the value observed by Kano & Tsuneta (1996, 1995) (Figure 4).

Porter & Klimchuk (1995) and Kankelborg *et al.* (1996, 1997) argue rather convincingly that there is no real correlation between a loop's observed length and maximum temperature. On the other hand, they argue that there is an inverse correlation between a loop's length and pressure. This seems to contradict the behavior of the quasi-static loop equation. A loop's length, $s_{max}$, is defined by $s_{max} \equiv s(T_{max})$, and $T_{max}$ is determined from $F_c(T_{max}) = 0$. The condition $F_c(T_{max}) = 0$ is met when $F_c^{\,2}(T_0)$ is exactly cancelled by $f_R$ minus $f_H$ in Equation (7). If a loop has a large pressure, we see that the term $f_H$ must grow more before it is able to overtake and cancel $f_R$. Hence, higher pressures should yield higher maximum temperatures and longer loops for a fixed conductive flux and heating input. On the other hand, by increasing the heating term and decreasing the conductive flux at the base we make $f_R$ minus $f_H$ more negative and also make the loop shorter with a lower maximum temperature. Hence, the way to create a situation in which a loop's pressure varies inversely with its length is for the heating rate to vary strongly with the inverse of the loop's length. Thus, the observed correlation between a loop's pressure and length seems to be due to a physical selection mechanism that has little to do with the form of the loop equation.

### *3.2 Heating the Lower Transition Region*

It is interesting to calculate the energy available for lower transition emission from our models. The total radiative flux from the lower transition region, including H Ly$\alpha$, is observed to be $5 \times 10^5$ ergs cm$^{-2}$ s$^{-1}$ (Timothy 1977, Vidal-Madjar 1977). Athay (1985) estimates the conductive energy flux parallel to magnetic field lines between $10^5$ K and $10^6$ K to be ~ $1 \times 10^6$ ergs cm$^{-2}$ s$^{-1}$. Using the conductive flux values from Table 1, the calculated disk coverages from Table 3, and correcting for active regions, we find our classical model (CL2) would conduct ~ $6 \times 10^5$ ergs cm$^{-2}$ s$^{-1}$ into the lower transition region. Our radiative model (R2) gives ~ $2 \times 10^4$ ergs cm$^{-2}$ s$^{-1}$. The classical type loops could provide, then, a significant fraction of the energy radiated as lower transition region emission in the quiet sun. If the conduction type loops (such as C2) were also present in large numbers, they could also provide a large fraction of the lower transition region's emission. With only a 30% coverage model C2 would conduct ~ $2 \times 10^5$ ergs cm$^{-2}$ s$^{-1}$into the lower transition region. On the other hand, the radiative output in the 170 – 175 Å bandpass would be a meager 70 erg ergs cm$^{-2}$ s$^{-1}$ (compare with the $3 \times 10^3$ ergs cm$^{-2}$ s$^{-1}$ for model CL2). The observational consequences of this are profound. It has been observed by several authors (Allen *et al.* 1997; Kankelborg *et al.* 1996, 1997; Brosius *et al.* 1996; Wang and Sheeley 1995) that the magnitude of the lower transition region emission at the footpoints of several coronal structures is consistent with the inferred conductive flux from the corona. It has also been observed that there are many fewer coronal structures than are necessary to account for the thousands of network elements observed at transition region temperatures (Feldman and Laming 1994). If the conductive type lukewarm loops were present in large numbers they would have modest emission in the upper transition region but footpoints in the lower transition region that would be as bright as the footpoints of structures with more prominent upper transition region emission. Our models could fit the observations of the solar atmosphere in a self-consistent manner providing a source for both the upper and lower transition region emissions.

We note that there is a large body of literature which argue against the concept that lower transition region emission is due to the interface of hotter plasmas with the chromosphere (for example, Feldman 1983, 1987, 1992, 1998; Feldman & Laming 1994; Dowdy *et al*. 1987). The focus of these "unresolved fine structure" models and "magnetic constriction" models is to show that there is a discontinuity between the properties of the corona and those of the transition region. They also argue that the magnetic structure of the solar atmosphere prevents significant conduction to the transition region from the corona. Hence, these authors conclude that the transition region from $3 \times 10^4$ K – $5 \times 10^5$ K is isolated from both the corona and the chromosphere. We will address these issues more directly in a later paper. For now we point out that our models represent plasmas at *sub*-coronal temperatures that are in thermal contact with the lower transition region and chromosphere. This study then casts the debate regarding the role of conduction in heating the solar transition region in a new light.

Several recent studies and observations directly support the notion that the interface of coronal and chromospheric plasmas provides a non-negligible contribution to lower transition region emission (Prés & Phillips 1999; Wikstøl *et al*. 1998; Oshea, Doyle, and Keenan 1998; Goodman 1998; Gallagher *et al*. 1998; Allen *et al*. 1997; Kankelborg *et al*. 1997, 1996; Ji, Song, and Hu 1996; Brosius *et al*. 1996; Wang and Sheeley 1995). We believe that these studies and our current study suggest a new model of the solar atmosphere in the quiet Sun. Our model is similar to the "magnetic junkyard" picture of Dowdy, Rabin, and Moore (1986), in that the distribution of structures in the solar atmosphere is determined by the distribution of magnetic elements in the supergranular network (and in the cell interiors as well, in our model). However, in light of the recent observations of Gallagher *et al*. (1998) who showed that network emission structures are approximately the same dimension from the lower transition region (30 000 K) to the upper transition region (400 000 K), we abandon the magnetic constriction of the Dowdy, Rabin, and Moore (1986) model. We also take into account the effects the dynamic nature the magnetic elements have in determining the properties of the plasma. In our picture, the upper transition region in the quiet sun is dominated by small-scale loops, for which the quasi-static description is valid; however, we cannot exclude the presence of other quasi-static structures such as funnels and dynamically evolving structures (such as jets, surges, plumes, flares, micro-flares, nano-flares). These quasi-static loop structures may be classical, conductive or radiative. We note that the context in which radiative type loops were observed was associated with active region transient phenomena; loops of this type may be important only for dynamically evolving structures. The quasi-static loops range in size from those recognized as bright points by Kankelborg *et al*. (L ~ 15 000 km) with $T_{max}$ ~ $1.5 \times 10^6$ K, to the micro-coronal bright points at the same temperatures analyzed by Falconer *et al*. (L ~ 3000 – 15 000 km), to the lukewarm loops modeled by Oluseyi *et al*. (L ~ 5000 km or smaller), with with $T_{max}$ ~ $7 \times 10^5$ K. In the Dowdy, Rabin and Moore model the emission throughout the transition region, at any particular temperature, is dominated by locally heated structures whose peak temperatures are at the temperature of interest. In our model, while the upper transition emission is dominated by structures whose peak temperatures are at upper transition region temperatures (*i.e.* lukewarm loops), conduction from coronal structures also contributes to upper transition region emission. In the lower transition region the emission in the quiet Sun is dominated primarily by the interface of hotter coronal and upper transition region plasmas with chromospheric plasmas. This model is attractive since it only requires a single heating mechanism from lower transition region to coronal temperatures. Antiochos & Noci point out that the observed form of the differential emission measure curve for the average Sun (averaged over space and time) appears to be universally obtained irrespective of the type of solar region one observes (e.g. Raymond & Doyle 19811, Withbroe 1981, Athay 1981), and is also appropriate for stars (Antiochos & Noci 1986). This observation, they point out, is consistent with the notion that the emission from all of these plasmas is due to a single mechanism.

Many other factors must be considered in developing a self-consistent model of the solar atmosphere. One is the contribution of dynamically evolving structures to transition region emission. Recent observations with high spatial and temporal resolutions indicate that dynamically evolving structures provide a non-negligible contribution to transition region emission (Korendyke *et al*. 1995, Yun *et al*. 1998, Kjeldseth-Moe & Brekke 1998, Pike & Mason 1998, Chae *et al*. 1998, Berghmans *et al.* 1998, Benz & Krucker 1998). Transition region emission lines have also been observed to show pervasive redshifts through several temperature regimes (Dere 1982, Gebbie *et al*. 1981). While our quasi-static models cannot directly address these dynamic observations, we note nonetheless that McClymont & Craig (1986, 1987) have developed models to explain the observed redshifts. The models they develop explain the observations by steady flows in cool loops ($T_c < 10^6$ K). Their models predict that loops with peak temperatures in the upper transition region can match the redshift observations only if they are short ($L < 10^9$ cm) with low pressures. These conditions are very similar to the conditions that characterize our conductive models.

If transition region loops are dynamic structures (and they probably are), we must ask ourselves for what portion of a loop's lifetime is the quasi-static model valid. Although any particular structure may not have its peak temperature in the transition region for very long, many may be evolving through that state. One may draw an analogy with ten year-olds in the human population. While ten year-olds are not a stable human configuration, if one observes humanity at any given time, one will always observe a significant population of ten year-olds. Likewise, a significant fraction of the transition region emission may be produced by structures evolving through the transition region.

Finally, one must also take account of the COmosphere postulated by Ayres and Rabin (1996). In their model the solar atmosphere has two components. One component exists only in association with magnetic flux tubes. This is the atmosphere where the temperature reversal occurs and includes the chromosphere, transition region and corona. The other component represents a hydrostatic atmosphere that does not undergo a temperature reversal. This component of the atmosphere is believed to be the source of the observed infrared emission lines of CO (Ayres and Rabin 1996). If the COmosphere exists, it would be radiatively heated by the hotter components of the atmosphere associated with magnetic flux tubes.

### *3.3 Conclusions*

We have shown that the assumptions of previous quasi-static loop models are not universally valid. In particular, the assumptions that are necessary to derive the RTV scaling law are shown to be invalid for short loops (< $10^9$ cm) at sub-coronal temperatures. We have found that three classes of solutions (classical, conductive and radiative) exist for quasi-static loop models matching our constraints. This is a new result. Loops that can be identified with each of these classes have been definitely observed (Kankelborg *et al.* 1996, 1997; Prés & Phillips 1999; Aschwanden *et al*. 1998). We note that one class of loops (the conductive loops) suggests the possible existence of a component of the solar atmosphere that could provide a significant amount of energy to cooler thermal layers of the solar atmosphere via thermal conduction, while not dominating the radiative output of plasmas at its peak temperature. These conduction type lukewarm loops may provide then an energy source for the observed lower transition region network elements that are not obviously associated with coronal structures. We have pointed out previously (Allen *et al*. 1996; Kankelborg *et al*. 1996, 1997) that the chromosphere/corona interface at the footpoints of polar plumes and XBPs may make a significant, perhaps dominant, contribution to the energy emitted by the local network in the lower transition region, contrary to the views expressed by a number of authors. In the present paper we show that it is possible that the interface between structures at sub-coronal temperatures and the chromosphere can make a significant contribution to the energy emitted by the lower transition region in the quiet sun.

## Table 1.

## Model Results

| Model Parameter | C1 | CL1 | R1 | C2 | CL2 | R2 | C3 | CL3 | R3 |
|---|---|---|---|---|---|---|---|---|---|
| $T_m$ ($\times 10^5$ K) | 5.89 | 5.62 | 6.02 | 7.41 | 7.59 | 7.08 | 8.91 | 8.91 | 8.71 |
| $\varepsilon$ ($\times 10^{-3}$ ergs cm$^{-3}$ s$^{-1}$) | 0.82 | 1.22 | 26.2 | 1.86 | 8.46 | 21.18 | 3.40 | 6.87 | 12.20 |
| $n_{e0}$ ($\times 10^9$ cm$^{-3}$) | 0.32 | 2.71 | 29.2 | 0.55 | 30.00 | 60.60 | 1.00 | 30.12 | 55.19 |
| $n_{eL}$ ($\times 10^9$ cm$^{-3}$) | 0.06 | 0.48 | 4.86 | 0.07 | 2.66 | 8.50 | 0.11 | 3.38 | 6.28 |
| L ($\times 10^8$ cm) | 5.00 | 4.29 | 1.27 | 5.00 | 3.21 | 2.16 | 5.27 | 4.69 | 4.66 |
| $F_{c0}$ ($\times 10^5$ ergs cm$^{-2}$ s$^{-1}$) | 2.72 | 2.74 | 2.70 | 6.20 | 9.00 | 2.27 | 11.50 | 11.67 | 5.39 |
| $F_R$ ($\times 10^5$ ergs cm$^{-2}$ s$^{-1}$) | 1.38 | 2.48 | 30.57 | 3.10 | 18.16 | 43.48 | 6.43 | 20.53 | 51.46 |
| $F_m = \varepsilon L$ ($\times 10^5$ ergs cm$^{-2}$ s$^{-1}$) | 4.10 | 5.22 | 33.27 | 9.30 | 27.16 | 45.75 | 17.93 | 32.2 | 56.85 |
| $F_R/F_c$ | 0.51 | 0.91 | ~ 11 | 0.50 | ~ 2 | ~ 20 | 0.56 | ~2 | ~10 |

| Model | $T_m$ ($\times 10^5$ K) | $\varepsilon$ ($\times 10^{-3}$ ergs cm$^{-3}$ s$^{-1}$) | $n_{e0}$ ($\times 10^9$ cm$^{-3}$) | $n_{eL}$ ($\times 10^9$ cm$^{-3}$) | L ($\times 10^8$ cm) | $F_{c0}$ ($\times 10^5$ ergs cm$^{-2}$ s$^{-1}$) |
|---|---|---|---|---|---|---|
| C1 | 5.89 | 0.82 | 0.32 | 0.06 | 5.00 | 2.72 |
| C2 | 7.41 | 1.86 | 0.55 | 0.07 | 5.00 | 6.20 |
| C3 | 8.91 | 3.40 | 1.00 | 0.11 | 5.27 | 11.50 |
| CL1 | 5.62 | 1.22 | 2.71 | 0.48 | 4.29 | 2.74 |
| CL2 | 7.59 | 8.46 | 30.00 | 2.66 | 3.21 | 9.00 |
| CL3 | 8.91 | 6.87 | 30.12 | 3.38 | 4.69 | 11.67 |
| R1 | 6.02 | 26.20 | 29.20 | 4.86 | 1.27 | 2.70 |
| R2 | 7.08 | 21.18 | 60.60 | 8.50 | 2.16 | 2.278 |
| R3 | 8.71 | 12.230 | 55.19 | 6.28 | 4.66 | 5.39 |

## Table 2.

## Line Emission for Loop Models

| Model | Fe IX (ergs/sec) | Fe X (ergs/sec) | O V (ergs/sec) | O VI (ergs/sec) | Ne IV (ergs/sec) | Ne V (ergs/sec) | Total (ergs/sec) |
|---|---|---|---|---|---|---|---|
| C1 | $9.34 \times 10^{15}$ | $5.12 \times 10^{14}$ | $3.09 \times 10^{15}$ | $7.61 \times 10^{15}$ | $9.90 \times 10^{14}$ | $5.14 \times 10^{14}$ | **$2.21 \times 10^{16}$** |
| C2 | $6.21 \times 10^{16}$ | $9.77 \times 10^{15}$ | $1.99 \times 10^{15}$ | $5.06 \times 10^{15}$ | $7.58 \times 10^{14}$ | $2.95 \times 10^{14}$ | **$8.00 \times 10^{16}$** |
| C3 | $2.52 \times 10^{17}$ | $7.16 \times 10^{16}$ | $9.05 \times 10^{15}$ | $1.37 \times 10^{16}$ | $3.58 \times 10^{15}$ | $8.13 \times 10^{14}$ | **$3.50 \times 10^{17}$** |
| CL1 | $3.50 \times 10^{17}$ | $1.51 \times 10^{16}$ | $1.96 \times 10^{17}$ | $2.94 \times 10^{17}$ | $7.81 \times 10^{16}$ | $2.30 \times 10^{16}$ | **$9.57 \times 10^{17}$** |
| CL2 | $4.63 \times 10^{19}$ | $7.31 \times 10^{18}$ | $5.57 \times 10^{18}$ | $6.88 \times 10^{18}$ | $2.23 \times 10^{18}$ | $4.68 \times 10^{17}$ | **$6.88 \times 10^{19}$** |
| CL3 | $1.87 \times 10^{20}$ | $4.97 \times 10^{19}$ | $1.17 \times 10^{19}$ | $1.51 \times 10^{19}$ | $4.65 \times 10^{18}$ | $9.81 \times 10^{17}$ | **$2.69 \times 10^{20}$** |
| R1 | $1.77 \times 10^{19}$ | $9.92 \times 10^{17}$ | $7.98 \times 10^{18}$ | $1.01 \times 10^{19}$ | $3.14 \times 10^{18}$ | $7.51 \times 10^{17}$ | **$4.07 \times 10^{19}$** |
| R2 | $2.03 \times 10^{20}$ | $2.14 \times 10^{19}$ | $2.62 \times 10^{19}$ | $4.42 \times 10^{19}$ | $1.03 \times 10^{19}$ | $2.90 \times 10^{18}$ | **$3.08 \times 10^{20}$** |
| R3 | $5.24 \times 10^{20}$ | $9.00 \times 10^{19}$ | $4.99 \times 10^{19}$ | $6.27 \times 10^{19}$ | $1.98 \times 10^{19}$ | $4.22 \times 10^{18}$ | **$7.51 \times 10^{20}$** |

## Table 3.

## Model Bandpass Luminosities

| Model | $T_m$ ($\times 10^5$ K) | Area Averaged Solar Bandpass Luminosity (ergs $cm^2$ $s^{-1}$) | % Coverage Necessary to Match Observational Data |
|---|---|---|---|
| C1 | 5.89 | $6.31 \times 10^{-1}$ | 100 |
| C2 | 7.41 | $2.29 \times 10^{0}$ | 100 |
| C3 | 8.91 | $9.49 \times 10^{0}$ | 100 |
| | | | |
| CL1 | 5.62 | $3.19 \times 10^{1}$ | 100 |
| CL2 | 7.59 | $3.06 \times 10^{3}$ | 65 – 80 |
| CL3 | 8.91 | $8.61 \times 10^{3}$ | 25 – 30 |
| | | | |
| R1 | 6.02 | $4.58 \times 10^{3}$ | 45 – 55 |
| R2 | 7.08 | $2.04 \times 10^{4}$ | ~ 10 |
| R3 | 8.71 | $2.30 \times 10^{4}$ | ~ 10 |

# Figure Legends

**Figure 1.** The corona as observed in the 171 –175 Å bandpass (Fe IX/X) with a Cassegrain multilayer telescope by Walker *et al.* (1988). This figure shows the diffusion emission on the disk that appears to have structures on the same scale as the supergranulation.

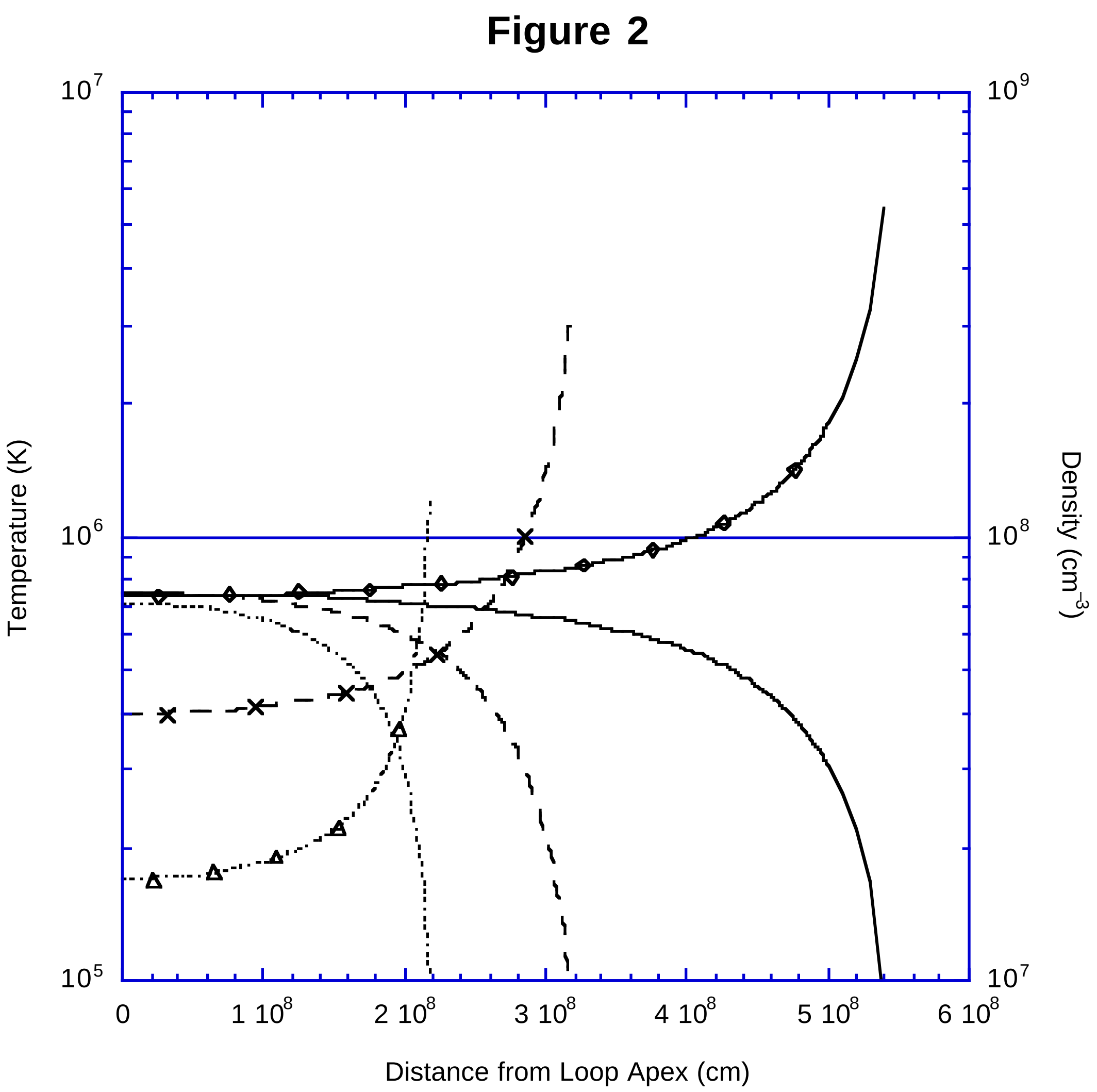

**Figure 2**. Temperature and density profiles calculated for representative loop models C2, CL2 and R2. Model C2 is represented by the solid lines, model CL2 by the dashed lines and model R2 by the dotted lines. The curves with markers are the density profiles for each respective model and the unmarked curves are the temperature profiles.

# Figure 3

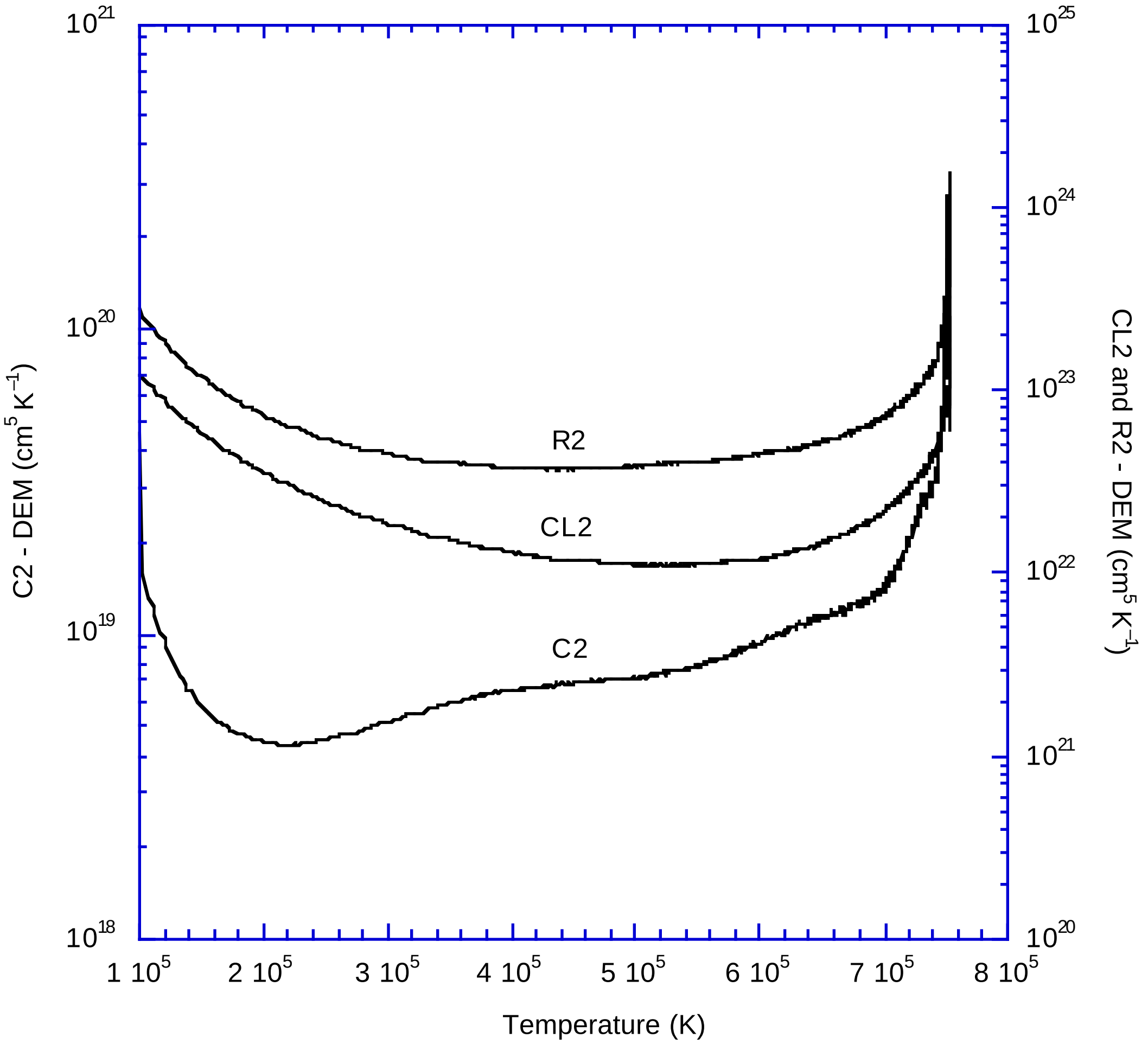

**Figure 3**. Differential emission measure curves derived for models C2, CL2 and R2.

# Figure 4

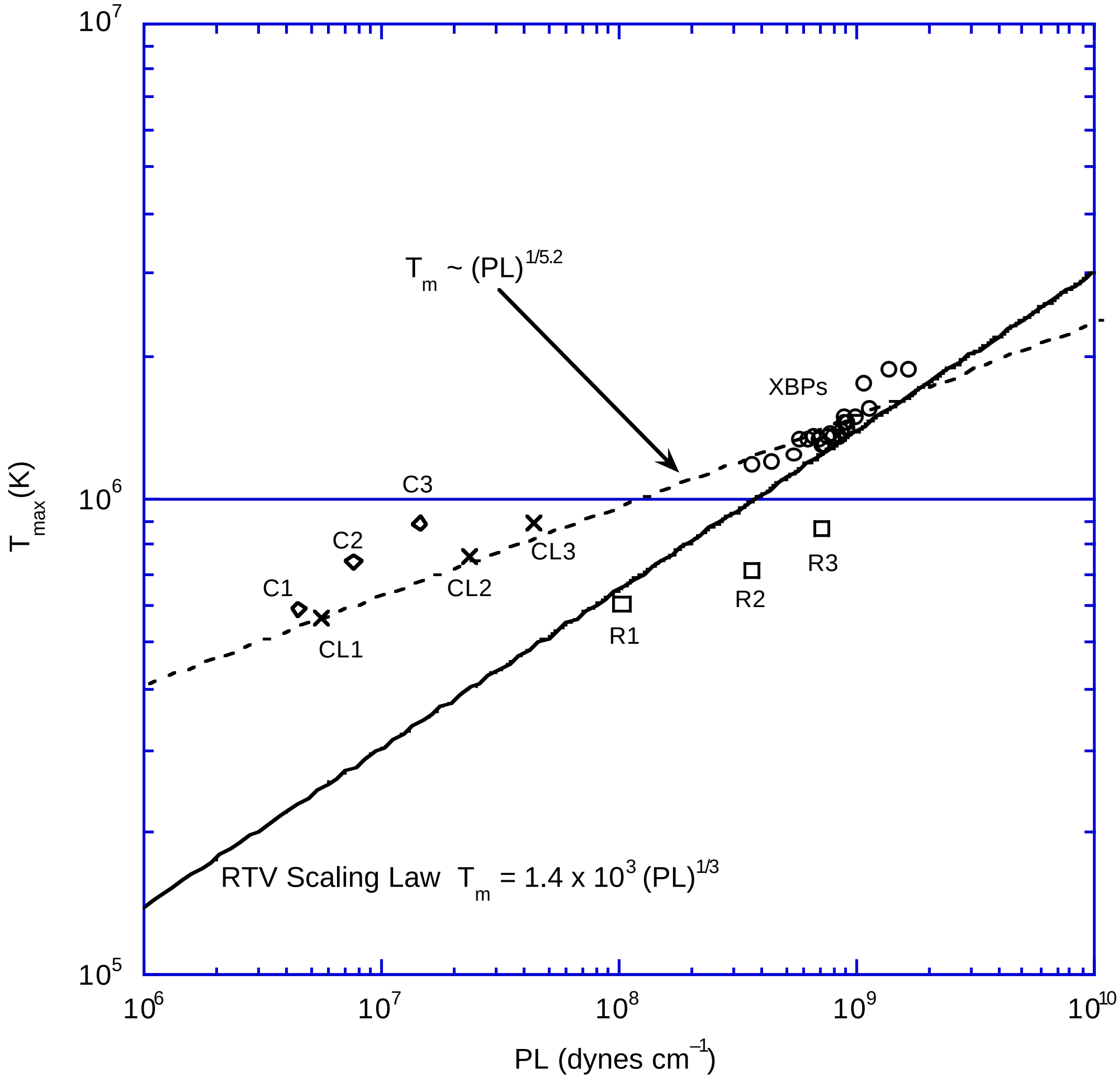

**Figure 4**. *T* vs. *PL* for our representative loop models, along with the 26 XBPs of Kankelborg *et al*.